\documentclass{jfm}
\usepackage{graphicx}
\usepackage{epstopdf, epsfig}
\usepackage{subfig}
\usepackage{amsmath}
\usepackage{amssymb}
\usepackage{amsfonts}
\usepackage{mathrsfs}
\usepackage{mathtools}
\usepackage{bm}
\usepackage{color}
\usepackage{array}
\usepackage{booktabs}
\usepackage{color}
\usepackage{siunitx}
\usepackage{gensymb}
\usepackage{multicol}
\usepackage{multirow}
\usepackage{ulem}
\allowdisplaybreaks[4]
\usepackage[T1]{fontenc}

\begin{document}
\shorttitle{Perturbation analysis of baroclinic torque}
\shortauthor{S. Zhang, Z. Xia, S. Chen}
\title{Perturbation analysis of baroclinic torque in low-Mach-number flows}

\author{Shengqi Zhang\aff{1}, Zhenhua Xia\aff{2}\corresp{\email{xiazh@zju.edu.cn}} and Shiyi Chen\aff{3,1}\corresp{\email{chensy@sustech.edu.cn}}}

\affiliation{\aff{1} State Key Laboratory for Turbulence and Complex Systems, Peking University, Beijing 100871, China\\
	\aff{2} Department of Engineering Mechanics, Zhejiang University, Hangzhou 310027, China\\
	\aff{3} Department of Mechanics and Aerospace Engineering, Southern University of Science and Technology, Shenzhen 518055, China}

\maketitle
\begin{abstract}
In this paper, we propse a series expansion of the baroclinic torque in low-Mach-number flows, so that the accuracy and universality of any buoyancy term could be examined analytically, and new types of buoyancy terms could be constructed and validated.
We first demonstrate that the purpose of introducing a buoyancy term is to approximate the baroclinic torque, and straightforwardly the accuracy of any buoyancy term could be measured by the deviation of its curl from the baroclinic torque.
Then a regular perturbation method is introduced for the elliptic equation of the hydrodynamic pressure in low-Mach-number flows, resulting in a sequence of Poisson equations, whose solutions lead to the series representation of the baroclinic torque and the new types of buoyancy terms. With the error definition of buoyancy terms and the series representation of the baroclinic torque,
the classical gravitational and centrifugal buoyancy term, as well as some other previously proposed buoyancy terms are revisited. 
Finally, numerical simulations confirm that, with a decreasing density variation or an increasing order of our newly proposed buoyancy term, the simplified equations with one of the new types of buoyancy terms can converge to the original low-Mach-number equations.
\end{abstract}

\section{Introduction}\label{sec:intro}

Buoyancy effect in variable-density flows is important especially for buoyancy-driven flows whose kinetic energy originates from the potential energy. Buoyancy effect is first modelled by the pioneering work from~\citet{boussinesq1903theorie}, who intuitively neglected small density fluctuations in the continuity equation and momentum equation except for the term combined with the gravitational acceleration $\bm{g}$, and introduced the gravitational buoyancy term $\rho'\bm{g}/\rho_0$. Here $\rho_0(t)\approx \rho(\bm{x},t)$ and $\rho'(\bm{x},t)=\rho(\bm{x},t)-\rho_0(t)$ are the constant reference density and the corresponding density fluctuation respectively. In recent decades, many theoretical and numerical results obtained using the Boussinesq's buoyancy term have shown good agreement with the experimental measurements, thereby validating the correctness and effectiveness of the Boussinesq's buoyancy term~\citep{sharp1983overview, ahlers2009heat, lohse2010small}. Besides the gravitational acceleration, the centrifugal acceleration in a rotating frame could also induce the buoyancy effect. This was modeled by~\citet{barcilon1967steady},~\citet{homsy1969centrifugally} and~\citet{busse1974convection} as $-\rho'\bm{\Omega}\times(\bm{\Omega}\times\bm{r})/\rho_0$, where $\bm{\Omega}$ is the angular velocity and $\bm{r}$ is the vector pointing from the origin on the rotating axis to the present location. For the flows where the different parts of the fluid container rotate independently,~\citet{lopez2013boussinesq} introduced the buoyancy term caused by the convection term $-\rho'\bm{u}\cdot\nabla\bm{u}/\rho_0$ to model the centrifugal buoyancy in non-rotating frame with local vortices.~\citet{kang2015thermal} further simplified Lopez's buoyancy term and applied it to a Taylor-Couette flow with variable density.~\citet{kang2019numerical} also introduced the buoyancy term $2\rho'\bm{u}\times\bm{\Omega}/\rho_0$ to account for the Coriolis effect.

The greatest advantage of the previous buoyancy terms is their simplicity, and thus many works applied the buoyancy terms in theoretical analysis and numerical simulations~\citep{sharp1983overview,ahlers2009heat,lohse2010small,chandrasekhar2013hydrodynamic,ng2015vertical,shishkina2016momentum,yang2016vertically,horn2018regimes}. However, among the buoyancy terms described above, only the classical gravitational buoyancy term~\citep{spiegel1960boussinesq,gray1976validity} and the classical centrifugal buoyancy term~\citep{barcilon1967steady} could be derived from the momentum equation for compressible flows.  The buoyancy terms introduced by~\citet{lopez2013boussinesq} and~\citet{kang2019numerical} are rather empirical and their validity remains uncertain. On the other hand, the accuracy of the previous buoyancy terms are very limited, and they can only be applied in flows with very small density variations. Furthermore, the invariance of previous buoyancy terms under frame transformations is questionable.


For variable-density flows with significant density variation, the incompressibility approximation with buoyancy terms listed above would lack in accuracy, and the low-Mach-number Navier-Stokes equations~\citep{paolucci1982filtering} should be used instead. By filtering out the weak sound waves and decomposing the pressure into a hydrodynamic pressure and a thermodynamic pressure, the low-Mach-number equations can overcome the difficulty in resolving sound speed ~\citep{paolucci1982filtering}.
Since they are more accurate than using the buoyancy terms when the flows with significant density variation are considered, many researchers applied the low-Mach-number equations in their numerical simulations~\citep{paolucci1990direct,livescu2007buoyancy,xia2016flow,livescu2020turbulence}. However, although the low-Mach-number approximation is accurate with Mach number approaching zero, the equation for the hydrodynamic pressure is no longer a Poisson equation, but a density coupled equation. This would greatly increase the difficulty in theoretical analysis. In addition, for the applications of the low-Mach-number equations in numerical simulations, some efficient numerical algorithms for the Poisson equation cannot be used and the equations have to be solved iteratively.


In a word, the previous buoyancy terms in the simplified momentum equation are simple but lack of the accuracy for flows with significant density variations. On the contrary, the low-Mach-number equations are accurate but complicated in theoretical analysis and numerical simulations. In the present paper, we would like to follow the traditional approach by introducing the buoyancy term in the modified momentum equation for flows with significant density variations. Our goal is to examine and increase the accuracy of the buoyancy terms.



\section{Analytic derivation}\label{sec:analytic}

\subsection{Low-Mach-number equations}\label{sec:analytic_LMN}
Considering a low-Mach-number variable-density flow with reference length scale $l_r^*$, reference velocity $u_r^*$, reference density $\rho_r^*$ and reference time scale $t_r^*=l_r^*/u_r^*$, the non-dimensionalized governing equations can be written as~\citep{paolucci1982filtering,majda1985derivation}:
\begin{subequations}
\begin{gather}
	\frac{1}{\rho}\frac{d\rho}{dt}
	+\nabla\cdot\bm{u}
	=0,
	\label{eqn:eqcontinuity}
	\\
	\frac{\partial\bm{u}}{\partial t}
	+\bm{u}\cdot\nabla\bm{u}
	=-\frac{1}{\rho}\nabla\Pi
	+\frac{1}{\rho}\nabla\cdot\mathbb{T}
	+\bm{f},
	\label{eqn:eqmomentum}
	\\
	\frac{\partial\rho}{\partial t}
	+\bm{u}\cdot\nabla\rho
	=Q.
	\label{eqn:eqdensity}
\end{gather}
\label{eqn:eqNS}
\end{subequations}
Here, the non-dimensionalized hydrodynamic pressure $\Pi=\Pi^*/(\rho_r^*u_r^{*2})$, viscous stress tensor $\mathbb{T}(\bm{x},t)=\mathbb{T}^*(\bm{x}^*,t^*)/(\rho_r^*u_r^{*2})$, and body force $\bm{f}(\bm{x},t)=\bm{f}^*(\bm{x}^*,t^*)/(l_r^{*-1}u_r^{*2})$. The expressions of $\mathbb{T}$ and $\bm{f}$ are assumed to be given. All effects contributing to the material derivative of $\rho$, such as heat conduction, mass diffusion, heat source/sink and chemical reactions~\citep{livescu2020turbulence}, are included in the right-hand side (R.H.S.) term $Q(\bm{x},t)=Q^*(\bm{x}^*,t^*)/(l_r^{*-1}\rho_r^*u_r^*)$ of equation~\eqref{eqn:eqdensity}, and $Q(\bm{x},t)$ is assumed to be a function of all possible variables except the velocity, such as $\bm{x}, t$, temperature, constituent concentration \textit{etc.}, and the governing equations for the variables contributing to $Q$ (temperature, constituent concentration \textit{etc.}) are assumed to be given.

In the present context, the fluid region $\mathcal{V}$ is assumed to be finite, simply connected and time-dependent with volume $V(t)$. The normal component of velocity at the boundary $\partial\mathcal{V}$ is assumed to be given at any time. All scalar, vector and tensor fields are assumed to be smooth. Integrating equation~\eqref{eqn:eqcontinuity} and~\eqref{eqn:eqdensity} in $\mathcal{V}$, the constraint of $Q$ for mass conservation can be derived according to the divergence theorem:
\begin{equation}
	\int_{\mathcal{V}}\frac{Q}{\rho}dV
	=\int_{\mathcal{V}}\frac{1}{\rho}\frac{d\rho}{dt}dV
	=-\int_{\mathcal{V}}\nabla\cdot\bm{u}dV
	=-\oint_{\partial\mathcal{V}}\bm{n}\cdot\bm{u}dS,
\end{equation}
Here $\bm{n}$ is the outwards-pointing unit normal vector at $\partial\mathcal{V}$.

Taking the divergence of equation~\eqref{eqn:eqmomentum} and projecting equation~\eqref{eqn:eqmomentum} to the normal direction at the boundary, we would obtain an elliptic equation of the hydrodynamic pressure $\Pi$ with Neumann boundary condition:
\begin{equation}
	\left\{
	\begin{aligned}
		\mathcal{V}:\quad&\nabla\cdot\left(\frac{1}{\rho}\nabla\Pi\right)
		=-\nabla\cdot\frac{\partial\bm{u}}{\partial t}
		-\nabla\cdot\left(\bm{u}\cdot\nabla\bm{u}\right)
		+\nabla\cdot\left(\frac{1}{\rho}\nabla\cdot\mathbb{T}\right)
		+\nabla\cdot\bm{f},
		\\
		\partial\mathcal{V}:\quad&\bm{n}\cdot\left(\frac{1}{\rho}\nabla\Pi\right)
		=-\bm{n}\cdot\frac{\partial\bm{u}}{\partial t}
		-\bm{n}\cdot\left(\bm{u}\cdot\nabla\bm{u}\right)
		+\bm{n}\cdot\left(\frac{1}{\rho}\nabla\cdot\mathbb{T}\right)
		+\bm{n}\cdot\bm{f}.
	\end{aligned}
	\right.
	\label{eqn:Pi}
\end{equation}
Although $\partial\bm{u}/\partial{t}$ is unknown until the momentum equation is solved, its divergence $\nabla\cdot(\partial\bm{u}/\partial{t})=-\partial(Q/\rho)/\partial{t}$ can be derived from equation~\eqref{eqn:eqcontinuity} and~\eqref{eqn:eqdensity}, and its wall-normal component $\bm{n}\cdot(\partial\bm{u}/\partial{t})=\partial(\bm{n}\cdot\bm{u})/\partial{t}-\bm{u}\cdot\partial\bm{n}/\partial{t}$ at the boundary $\partial\mathcal{V}$ is determined by the velocity boundary condition and the motion of $\mathcal{V}$.

For simplicity, we introduce the following denotations:
$$\bm{A}=-\frac{\partial\bm{u}}{\partial t},\quad \bm{C}=-\bm{u}\cdot\nabla\bm{u},\quad \mathrm{and } \quad\bm{D}=\rho^{-1}\nabla\cdot\mathbb{T}.$$
Straightforwardly, the hydrodynamic pressure $\Pi$ could be separated as
\begin{equation}
	\Pi=\Pi_{\bm{A}}+\Pi_{\bm{C}}+\Pi_{\bm{D}}+\Pi_{\bm{f}},
	\label{eqn:Pi_divide}
\end{equation}
and the four corresponding elliptic equations and Neumann boundary conditions are of the same form:
\begin{equation}
\left\{
\begin{aligned}
	\mathcal{V}:\quad&\nabla\cdot\left(\frac{1}{\rho}\nabla\Pi_{\bm{Y}}\right)
	=\nabla\cdot\bm{Y},
	\\
	\partial\mathcal{V}:\quad&\bm{n}\cdot\left(\frac{1}{\rho}\nabla\Pi_{\bm{Y}}\right)
	=\bm{n}\cdot\bm{Y},
\end{aligned}
\right.
\label{eqn:Pi_Y}
\end{equation}
with $\bm{Y}\in\left\{\bm{A},\bm{C},\bm{D},\bm{f}\right\}$. For example, for $\Pi_{\bm{A}}$, the governing equation and the corresponding boundary conditions are
\begin{equation}
\left\{
\begin{aligned}
\mathcal{V}:\quad &\nabla\cdot\left(\frac{1}{\rho}\nabla\Pi_{\bm{A}}\right)
		=\nabla\cdot\left(-\frac{\partial\bm{u}}{\partial t}\right)
		=\frac{\partial}{\partial t}\left(\frac{Q}{\rho}\right),
		\\
		\partial\mathcal{V}:\quad &\bm{n}\cdot\left(\frac{1}{\rho}\nabla\Pi_{\bm{A}}\right)
		=\bm{n}\cdot\left(-\frac{\partial\bm{u}}{\partial t}\right)=-\frac{\partial(\bm{n}\cdot\bm{u})}{\partial{t}}+\frac{\partial\bm{n}}{\partial{t}}\cdot\bm{u}
\end{aligned}
\right.
\label{eqn:Pi_A}
\end{equation}

\subsection{Baroclinic torque}\label{sec:analytic_vorticity}
The curl of equation~\eqref{eqn:eqmomentum} gives the equation of vorticity $\bm{\omega}=\nabla\times\bm{u}$
\begin{equation}
\frac{\partial\bm{\omega}}{\partial t}
	=\frac{1}{\rho^2}\nabla\rho\times\nabla\Pi
	+\nabla\times\left(-\bm{u}\cdot\nabla\bm{u}
	+\frac{1}{\rho}\nabla\cdot\mathbb{T}
	+\bm{f}\right).
\end{equation}
The contribution $\rho^{-2}\nabla\rho\times\nabla\Pi$ to $\partial\bm{\omega}/\partial t$ is called baroclinic torque~\citep{mcmurtry1989effects,jahanbakhshi2015baroclinic}. Following the decomposition of hydrodynamic pressure~\eqref{eqn:Pi_divide}, the baroclinic torque could also be decomposed into four parts
\begin{equation}
	\frac{1}{\rho^2}\nabla\rho\times\nabla\Pi
	=\frac{1}{\rho^2}\nabla\rho\times\nabla\Pi_{\bm{A}}
	+\frac{1}{\rho^2}\nabla\rho\times\nabla\Pi_{\bm{C}}
	+\frac{1}{\rho^2}\nabla\rho\times\nabla\Pi_{\bm{D}}
	+\frac{1}{\rho^2}\nabla\rho\times\nabla\Pi_{\bm{f}},
\end{equation}
with the four terms in the R.H.S. corresponding to $\bm{A},\bm{C},\bm{D},\bm{f}$ respectively. Therefore, we may arrive at the following equation
\begin{align}
	\frac{\partial\bm{\omega}}{\partial t}&=\frac{1}{\rho^2}\nabla\rho\times\nabla\left(\Pi_{\bm{A}}+\Pi_{\bm{C}}+\Pi_{\bm{D}}+\Pi_{\bm{f}}\right)
	+\nabla\times\left(\bm{C}+\bm{D}+\bm{f}\right)
	\notag \\
	&=\frac{1}{\rho^2}\nabla\rho\times\nabla\Pi_{\bm{A}}
	+\left(\nabla\times\bm{C}+\frac{1}{\rho^2}\nabla\rho\times\nabla\Pi_{\bm{C}}\right)
	\label{eqn:omega} \\
	&\qquad\qquad\qquad\ \ \ \ +\left(\nabla\times\bm{D}+\frac{1}{\rho^2}\nabla\rho\times\nabla\Pi_{\bm{D}}\right)
	+\left(\nabla\times\bm{f}+\frac{1}{\rho^2}\nabla\rho\times\nabla\Pi_{\bm{f}}\right).
	\notag
\end{align}
Obviously, the actual contribution of any term $\bm{Y}\in\left\{\bm{C},\bm{D},\bm{f}\right\}$ to the vorticity evolution $\partial\bm{\omega}/\partial t$ should be
\begin{equation}
	\nabla\times\bm{Y}+\frac{1}{\rho^2}\nabla\rho\times\nabla\Pi_{\bm{Y}}.
	\label{eqn:F_omega}
\end{equation}
The term
\begin{equation}
	\frac{1}{\rho^2}\nabla\rho\times\nabla\Pi_{\bm{A}},
	\label{eqn:A_omega}
\end{equation}
denotes the joint contributions from $\partial (\rho^{-1}Q)/\partial t$ and $\left(\bm{n}\cdot\partial\bm{u}/\partial t\right)|_{\partial\mathcal{V}}$. Please note that the gravity $\bm{g}$ and the centrifugal force $-\bm{\Omega}\times(\bm{\Omega}\times\bm{r})$ are both irrotational fields with zero curl, they could only drive the flows through the corresponding partial baroclinic torque $\rho^{-2}\nabla\rho\times\nabla\Pi_{\bm{f}}$.

For a better demonstration of the magnitude of baroclinic torque, let's define $\rho_0(t)>0$ as the time-dependent reference density (different from the constant $\rho_r$) with the corresponding density fluctuation $\rho'(\bm{x},t)=\rho-\rho_0$. Consequently the baroclinic torque corresponding to $\bm{Y}\in\left\{\bm{A},\bm{C},\bm{D},\bm{f}\right\}$ could be rewritten as {\iffalse}(see the supplementary material for detailed derivation){\fi}
\begin{equation}\begin{aligned}
	\frac{1}{\rho^2}\nabla\rho\times\nabla\Pi_{\bm{Y}}
	&=\nabla\times\left(\frac{\rho'}{\rho_0\rho}\nabla\Pi_{\bm{Y}}\right)
	=\frac{1}{\rho_0}\nabla\left(\frac{\rho'}{\rho}\right)\times\nabla\Pi_{\bm{Y}}
	\label{eqn:baro_eps_eta_Y}
\end{aligned}\end{equation}
and the total baroclinic torque could be written as
\begin{align}
	\frac{1}{\rho^2}\nabla\rho\times\nabla\Pi
	=\frac{1}{\rho_0}\nabla\left(\frac{\rho'}{\rho}\right)\times\nabla\Pi
	\label{eqn:baro_eps_eta}
\end{align}

\subsection{Buoyancy term}\label{sec:analytic_buoyancy}

It was shown in \S\ref{sec:analytic_vorticity} that the buoyancy effect caused by gravity or centrifugal force originates from the baroclinic torque. Here we would further clarify the purpose of introducing a buoyancy term. That is, with a buoyancy term $\bm{B}$ adopted, the following modified equations
\begin{subequations}
	\begin{gather}
		\frac{1}{\rho}\frac{d\rho}{dt}
		+\nabla\cdot\bm{u}
		=0,
		\label{eqn:eqcontinuity_buoy}
		\\
		\frac{\partial\bm{u}}{\partial t}
		+\bm{u}\cdot\nabla\bm{u}
		=-\frac{1}{\rho_0}\nabla\widetilde{\Pi}
		+\frac{1}{\rho}\nabla\cdot\mathbb{T}
		+\bm{f}
		+\bm{B},
		\label{eqn:eqmomentum_buoy}
		\\
		\frac{\partial\rho}{\partial t}
		+\bm{u}\cdot\nabla\rho
		=Q,
		\label{eqn:eqdensity_buoy}
	\end{gather}
	\label{eqn:eqNS_buoy}
\end{subequations}
should be equivalent to equations~\eqref{eqn:eqNS}. Here, $\widetilde{\Pi}$ is the new hydrodynamic pressure and it satisfies a Poisson equation with compatible Neumann boundary condition:
\begin{equation}
	\left\{
	\begin{aligned}
		\mathcal{V}:\quad&\frac{1}{\rho_0}\nabla^2\widetilde{\Pi}
		=\nabla\cdot\left(-\frac{\partial\bm{u}}{\partial t}
		-\bm{u}\cdot\nabla\bm{u}
		+\frac{1}{\rho}\nabla\cdot\mathbb{T}
		+\bm{f}
		+\bm{B}\right),
		\\
		\partial\mathcal{V}:\quad&\frac{1}{\rho_0}\bm{n}\cdot\nabla\widetilde{\Pi}
		=\bm{n}\cdot\left(-\frac{\partial\bm{u}}{\partial t}
		-\bm{u}\cdot\nabla\bm{u}
		+\frac{1}{\rho}\nabla\cdot\mathbb{T}
		+\bm{f}
		+\bm{B}\right).
	\end{aligned}
	\right.
	\label{eqn:Pi_buoy}
\end{equation}
Since $\nabla\times(\nabla\widetilde{\Pi}/\rho_0)=0$, the baroclinic torque disappeared in the vorticity equation corresponding to equation~\eqref{eqn:eqmomentum_buoy}:
\begin{equation}
	\frac{\partial\bm{\omega}}{\partial t}
	=\nabla\times\bm{B}
	+\nabla\times\left(-\bm{u}\cdot\nabla\bm{u}
	+\frac{1}{\rho}\nabla\cdot\mathbb{T}
	+\bm{f}\right).
	\label{eqn:omega_buoy}
\end{equation}

For a flow problem with given $\bm{u}|_{\partial\mathcal{V}}$, $\bm{u}$ could be uniquely determined  with known $\nabla\cdot\bm{u}$ and $\nabla\times\bm{u}$. For the present flow problem, the governing equation for velocity divergence, $\partial(\nabla\cdot\bm{u})/\partial{t}=-\partial\left(\rho^{-1}Q\right)/\partial{t}$, is kept the same as the original one, and the governing equation~\eqref{eqn:omega_buoy} of $\bm{\omega}$ should be the same as the original vorticity equation~\eqref{eqn:omega} if equations~\eqref{eqn:eqNS_buoy} are equivalent to the original low-Mach-number equations~\eqref{eqn:eqNS}. Therefore, the buoyancy term $\bm{B}$ should satisfy the constraint that
\begin{equation}~\label{eqn:B-pi}
\nabla\times\bm{B}=\frac{1}{\rho^2}\nabla\rho\times\nabla\Pi.
\end{equation}
On the other hand, if the constraint~\eqref{eqn:B-pi} is satisfied, the velocity field obtained from equations~\eqref{eqn:eqNS_buoy} is equal to that from the original low-Mach-number equations~\eqref{eqn:eqNS}. In other words, the purpose of introducing a buoyancy term is to approximate the baroclinic torque, and the curl of the buoyancy term in the modified momentum equation~\eqref{eqn:eqmomentum_buoy} should compensate the loss of baroclinic torque in the original low-Mach-number equations, so that equations~\eqref{eqn:eqNS_buoy} are equivalent to equations~\eqref{eqn:eqNS}.

For an arbitrary buoyancy term, if the constraint~\eqref{eqn:B-pi} was not satisfied accurately, we could introduce the terms
\begin{equation}
	\left\|\nabla\times\bm{B}-\frac{1}{\rho^2}\nabla\rho\times\nabla\Pi\right\|
	\ \text{and}\
	\left.\left\|\nabla\times\bm{B}-\frac{1}{\rho^2}\nabla\rho\times\nabla\Pi\right\|\middle/\left\|\frac{1}{\rho^2}\nabla\rho\times\nabla\Pi\right\|\right..
	\label{eqn:Err_buoy}
\end{equation}
to characterize its error and relative error. Here, the spatial $L-2$ norms $\|\cdot\|$ for vector and scalar fields are defined as
\begin{equation}
	\|\bm{v}\|
	=\sqrt{\frac{1}{V}\int_{\mathcal{V}}{\bm{v}\cdot\bm{v}dV}}
	,\quad
	\|\phi\|
	=\sqrt{\frac{1}{V}\int_{\mathcal{V}}{\phi^2dV}}.
	\label{eqn:L2norm}
\end{equation}



Following the decomposition of hydrodynamic pressure (equation~\eqref{eqn:Pi_divide}), a buoyancy term $\bm{B}$ could also be decomposed into four partial buoyancy terms corresponding to $\bm{A},\bm{C},\bm{D},\bm{f}$ respectively:
\begin{equation}
	\bm{B}=\bm{B}_{\bm{A}}+\bm{B}_{\bm{C}}+\bm{B}_{\bm{D}}+\bm{B}_{\bm{f}}.
	\label{eqn:B_divide}
\end{equation}
That is, the target of a partial buoyancy term $\bm{B}_{\bm{Y}}$ with $\bm{Y}\in\{\bm{A},\bm{C},\bm{D},\bm{f}\}$, is to approximate the corresponding partial baroclinic torque $\rho^{-2}\nabla\rho\times\nabla\Pi_{\bm{Y}}$. Therefore the error and relative error of $\bm{B}_{\bm{Y}}$ can be defined respectively as
\begin{equation}
	\left\|\nabla\times\bm{B}_{\bm{Y}}-\frac{1}{\rho^2}\nabla\rho\times\nabla\Pi_{\bm{Y}}\right\|
	\ \text{and}\
	\left.\left\|\nabla\times\bm{B}_{\bm{Y}}-\frac{1}{\rho^2}\nabla\rho\times\nabla\Pi_{\bm{Y}}\right\|\middle/\left\|\frac{1}{\rho^2}\nabla\rho\times\nabla\Pi_{\bm{Y}}\right\|\right..
	\label{eqn:Err_buoy_Y}
\end{equation}

It should be noted that the above decomposition of a buoyancy term~\eqref{eqn:B_divide} is non-unique since we only require that the buoyancy term $\bm{B}$ satisfies the constraint~\eqref{eqn:B-pi}. Nevertheless, a preferred form could be chosen for convenience. Inspired by the derivation of equation~\eqref{eqn:baro_eps_eta_Y}, an ``exact'' partial buoyancy term corresponding to $\bm{Y}\in\{\bm{A},\bm{C},\bm{D},\bm{f}\}$ could be written as
\begin{align}
\bm{B}^{\dagger}_{\bm{Y}}
=\frac{\rho'}{\rho_0\rho}\nabla\Pi_{\bm{Y}},
\end{align}
and the ``exact'' buoyancy term could be written as
\begin{align}
 	\bm{B}^{\dagger}
 	=\frac{\rho'}{\rho_0\rho}\nabla\Pi
 	=\bm{B}^{\dagger}_{\bm{A}}+\bm{B}^{\dagger}_{\bm{C}}+\bm{B}^{\dagger}_{\bm{D}}+\bm{B}^{\dagger}_{\bm{f}}.
\end{align}



\subsection{Regular perturbation method for the hydrodynamic pressure}\label{sec:analytic_perturbation}


In order to obtain the error scaling of any buoyancy term with respect to the amplitude of density variation, a proper expansion of the baroclinic torque should be derived. This could be achieved by introducing a regular perturbation method for the equation of hydrodynamic pressure. With the previous definition of $\rho_0$ and $\rho'$ in \S\ref{sec:analytic_vorticity}, equation~\eqref{eqn:Pi_Y} could be rewritten as
\begin{equation}
	\left\{
	\begin{aligned}
		\mathcal{V}:\quad&\nabla\cdot\left[\left(1-\frac{\rho'}{\rho}\right)\nabla\Pi_{\bm{Y}}\right]
		=\nabla\cdot\left(\rho_0\bm{Y}\right),
		\\
		\partial\mathcal{V}:\quad&\bm{n}\cdot\left[\left(1-\frac{\rho'}{\rho}\right)\nabla\Pi_{\bm{Y}}\right]
		=\bm{n}\cdot\left(\rho_0\bm{Y}\right).
	\end{aligned}
	\right.
	\label{eqn:Pi_Y_rhop}
\end{equation}
since $\rho_0$ depends only on $t$. In addition, we define $\epsilon(t)=\max_{\mathcal{V}}\left[|\rho'/\rho|\right]$ as the maximum value of relative density fluctuation, and define $\eta(\bm{x},t)=(\rho'/\rho)/\epsilon(t)$ whose absolute value is always no larger than $1$. If $\epsilon=0$, equation~{\eqref{eqn:Pi_Y_rhop}} is exactly the Poisson equation. If the fluid has density variation ($\epsilon>0$), we could choose an ansatz for the expansion of $\nabla\Pi_{\bm{Y}}$ with respect to $\epsilon$
\begin{equation}
\begin{aligned} \nabla\Pi_{\bm{Y}}&=\nabla\Pi_{\bm{Y}}^{(0)}+\epsilon\nabla\Pi_{\bm{Y}}^{(1)}+\epsilon^2\nabla\Pi_{\bm{Y}}^{(2)}+\cdots
	\\
	&=\sum_{k=0}^{\infty}{\left(\epsilon^k\nabla\Pi_{\bm{Y}}^{(k)}\right)}.
\end{aligned}
\label{eqn:Pi_ansatz}
\end{equation}

By taking ansatz~\eqref{eqn:Pi_ansatz} into equation~\eqref{eqn:Pi_Y_rhop}, $\nabla\Pi_{\bm{Y}}^{(k)}$ in~\eqref{eqn:Pi_ansatz} could be computed from the corresponding Poisson equation led by $\epsilon^k$ for each $k\geq0$ (notice that $\rho'/\rho=\epsilon\eta$)
\begin{equation}
\begin{aligned}
	\epsilon^0:\quad&\left\{
	\begin{aligned}
		\mathcal{V}:\quad&\nabla^2\Pi_{\bm{Y}}^{(0)}
		=\nabla\cdot\left(\rho_0\bm{Y}\right),
		\\
		\partial\mathcal{V}:\quad&\bm{n}\cdot\nabla\Pi_{\bm{Y}}^{(0)}
		=\bm{n}\cdot\left(\rho_0\bm{Y}\right),
	\end{aligned}
	\right.
	\\
	\epsilon^k,\ k\geq1:\quad&\left\{
	\begin{aligned}
		\mathcal{V}:\quad&\epsilon^k\nabla^2\Pi_{\bm{Y}}^{(k)}
		=\epsilon^{k-1}\nabla\cdot\left(\frac{\rho'}{\rho}\nabla\Pi_{\bm{Y}}^{(k-1)}\right),
		\\
		\partial\mathcal{V}:\quad&\epsilon^k\bm{n}\cdot\nabla\Pi_{\bm{Y}}^{(k)}
		=\epsilon^{k-1}\bm{n}\cdot\left(\frac{\rho'}{\rho}\nabla\Pi_{\bm{Y}}^{(k-1)}\right).
	\end{aligned}
	\right.
\end{aligned}
\label{eqn:Pi_ansatz_poisson}
\end{equation}

It can be proved that {\iffalse}(see the supplementary material){\fi}
\begin{align}
	\sum_{k=0}^{\infty}{\left(\epsilon^k\nabla\Pi_{\bm{Y}}^{(k)}\right)}
	=\nabla\Pi_{\bm{Y}},
	\label{eqn:Pi_Y_ansatz_proof}
\end{align}
when $\epsilon<1$, i.e. the ansatz~\eqref{eqn:Pi_ansatz} is correct when $\max_{\mathcal{V}}\left[|\rho'/\rho|\right]<1$.

With the expansion~\eqref{eqn:Pi_ansatz}, the partial baroclinic torque and the ``exact'' partial buoyancy term corresponding to $\bm{Y}\in\left\{\bm{A},\bm{C},\bm{D},\bm{f}\right\}$ could be expanded when $\epsilon<1$:
\begin{align}
	\frac{1}{\rho^2}\nabla\rho\times\nabla\Pi_{\bm{Y}}
	=\frac{1}{\rho_0}\nabla\left(\frac{\rho'}{\rho}\right)\times\sum_{k=0}^{\infty}{\left(\epsilon^k\nabla\Pi_{\bm{Y}}^{(k)}\right)}
	,\quad
	\bm{B}_{\bm{Y}}^{\dagger}
	=\frac{\rho'}{\rho_0\rho}\sum_{k=0}^{\infty}{\left(\epsilon^k\nabla\Pi_{\bm{Y}}^{(k)}\right)}.
	\label{eqn:baro_buoy_expansion}
\end{align}
The $n^{\text{th}}$-order partial buoyancy terms could be defined with the leading terms of the ``exact'' partial buoyancy terms:
\begin{align}
	\bm{B}_{\bm{Y}}^{(n)}=\frac{\rho'}{\rho_0\rho}\sum_{k=0}^{n-1}{\left(\epsilon^k\nabla\Pi_{\bm{Y}}^{(k)}\right)},\quad n\geq1.
	\label{eqn:buoy_new}
\end{align}

Therefore, we can define the $n^{\text{th}}$-order type-I buoyancy terms straightforwardly as
\begin{align}
	n\geq1:\ \bm{B}^{(n)}=\bm{B}_{\bm{A}}^{(n)}+\bm{B}_{\bm{C}}^{(n)}+\bm{B}_{\bm{D}}^{(n)}+\bm{B}_{\bm{f}}^{(n)}.
\label{eqn:T1buoy}
\end{align}
{\iffalse}The detailed computation procedure is shown in the supplementary material.{\fi}

In some circumstances, the flow is driven by the buoyancy effect induced by a large conservative body force $\bm{f}=\nabla\Phi$. According to equation~\eqref{eqn:Pi_ansatz_poisson}, there is $\nabla\Pi_{\bm{f}}^{(0)}=\rho_0\nabla\Phi_g=\rho_0\bm{f}$. For this special kind of flows, we can define the $n^{\text{th}}$-order type-II buoyancy terms as
\begin{equation}\begin{aligned}
	n=1:\ \hat{\bm{B}}^{(1)}&=\frac{\rho'}{\rho_0\rho}\nabla\Pi_{\bm{f}}^{(0)}=\frac{\rho'}{\rho}\bm{f},
	\\
	n\geq2:\ \hat{\bm{B}}^{(n)}&=\bm{B}^{(n-1)}_{\bm{A}}+\bm{B}^{(n-1)}_{\bm{C}}+\bm{B}^{(n-1)}_{\bm{D}}+\bm{B}^{(n)}_{\bm{f}}.
\end{aligned}\label{eqn:T2buoy}\end{equation}
{\iffalse}The detailed computation procedure is also shown in the supplementary material.{\fi}

For type-I and type-II buoyancy terms, the ``$n^{\text{th}}$-order'' only denotes the highest order of expansion, instead of the accuracy. It should be noticed that,{\iffalse}supplementary{\fi} when $\bm{f}$ is conservative, the $n^{\text{th}}$-order type-I buoyancy term requires to solve $n$ Poisson equations, while the $n^{\text{th}}$-order type-II buoyancy term only requires to solve $(n-1)$ Poisson equations. In the following, we will show that the relative errors for both types of $n^{\text{th}}$-order buoyancy terms are $O(\epsilon^n)$ convergent.

\subsection{Accuracy of buoyancy terms}\label{sec:analytic_accuracy}
In order to examine the accuracy of buoyancy terms at a fixed time $t_0$, a continuous set of instantaneous flow fields and reference densities {$\left\{\left(\bm{u}(\bm{x},t_0;\lambda),\rho(\bm{x},t_0;\lambda),\rho_0(t_0;\lambda)\right)\right\}$} with a continuous parameter $\lambda\in(0,\infty)$ is defined. In addition, we assume that $\bm{u}(\bm{x},t_0;\lambda)$ and $\rho(\bm{x},t_0;\lambda)$ are uniformly continuous in $(\bm{x},\lambda)$ space, $\rho_0(t_0;\lambda)$ and $\epsilon(\lambda)=\max_{\mathcal{V}}\left[|(\rho-\rho_0)/\rho|\right]$ are continuous with $\lambda$, and that
\begin{align}
	\lim\limits_{\lambda\rightarrow0^+}{\epsilon}=0.
	\label{eqn:eps_converge_0}
\end{align}
Naturally, a scalar $\phi(\lambda)$ is acknowledged to have $O(\epsilon^{\alpha})$ scaling with $\alpha\in\mathbb{R}$, if
\begin{align}
	\lim\limits_{\lambda\rightarrow0^+}{\frac{\phi}{\epsilon^{\alpha}}}=A,\quad 0<|A|<\infty.
	\label{eqn:On_define}
\end{align}

It should be emphasized that, a valid buoyancy term must at least have a relative error tending to $0$ when $\epsilon\rightarrow0$. This is because when compared with a trivial buoyancy term $0$, a buoyancy term with $O(1)$ relative error may cause a larger deviation from the real physics, even for very small $\epsilon$.

With the definitions of error and relative error of a buoyancy term~\eqref{eqn:Err_buoy} or a partial buoyancy term~\eqref{eqn:Err_buoy_Y} introduced in \S\ref{sec:analytic_buoyancy}, and the perturbation solution~\eqref{eqn:baro_buoy_expansion} of baroclinic torque in \S\ref{sec:analytic_perturbation}, the accuracy of buoyancy terms could be examined:

\subsubsection{Classical gravitational buoyancy term $\rho'\bm{g}/\rho_0$}
The classical gravitational buoyancy term $\rho'\bm{g}/\rho_0$ is introduced by~\citet{boussinesq1903theorie}. Here we only consider Newtonian gravity, so the corresponding body force $\bm{f}=\bm{g}=\nabla\Phi_g$ is conservative. According to equation~\eqref{eqn:Pi_ansatz_poisson}, there is
\begin{align}
	\nabla\Pi_{\bm{f}}^{(0)}=\rho_0\nabla\Phi_g=\rho_0\bm{f}.
	\label{eqn:Phi_Pi0}
\end{align}
It could be proved {\iffalse}(see the supplementary material) {\fi}that the relative error of $\bm{B}_{\bm{f}}=\rho'\bm{g}/\rho_0$ is $O(\epsilon)$ convergent in general cases:
\begin{align}
	\lim\limits_{\lambda\rightarrow0^+}{\left\{\left.\frac{\left\|\nabla\times\left(\rho'\rho_0^{-1}\bm{g}\right)-\rho^{-2}\nabla\rho\times\nabla\Pi_{\bm{f}}\right\|}{\left\|\rho^{-2}\nabla\rho\times\nabla\Pi_{\bm{f}}\right\|}\middle/\epsilon\right.\right\}}
	=\lim\limits_{\lambda\rightarrow0^+}\frac{\left\|\nabla\eta\times\left(2\eta\nabla\Pi_{\bm{f}}^{(0)}-\nabla\Pi_{\bm{f}}^{(1)}\right)\right\|}{\left\|\nabla\eta\times\nabla\Pi_{\bm{f}}^{(0)}\right\|}.
	\label{eqn:Err_buoy_g}
\end{align}
Therefore the classical gravitational buoyancy term is valid, with relative error tending to $0$ when $\epsilon\rightarrow0$. Equation~\eqref{eqn:Err_buoy_g} could help to decide whether the classical gravitational buoyancy term is enough for a required accuracy.

\subsubsection{Classical centrifugal buoyancy term $-\rho'\bm{\Omega}\times(\bm{\Omega}\times\bm{r})/\rho_0$}
The classical centrifugal buoyancy term $-\rho'\bm{\Omega}\times(\bm{\Omega}\times\bm{r})/\rho_0$ is introduced by~\citet{barcilon1967steady, homsy1969centrifugally, busse1974convection}. For this buoyancy term, the corresponding body force $\bm{f}=-\bm{\Omega}\times(\bm{\Omega}\times\bm{r})=\nabla\Phi_c$ is conservative. If we use cylindrical coordinate with $z$ axis being the rotating axis, there is $\bm{f}=\Omega^2r\bm{e}_r$ and $\Phi_c=\Omega^2r^2/2$. According to equation~\eqref{eqn:Pi_ansatz_poisson}, there is
\begin{align}
	\nabla\Pi_{\bm{f}}^{(0)}=\rho_0\nabla\Phi_c=\rho_0\bm{f}.
	\label{eqn:Phi_Pi0_cent}
\end{align}
It could also be proved {\iffalse}(see the supplementary material) {\fi}that $\bm{B}_{\bm{f}}=-\rho'\bm{\Omega}\times(\bm{\Omega}\times\bm{r})/\rho_0$ has an $O(\epsilon)$ relative error in general cases:
\begin{align}
	\lim\limits_{\lambda\rightarrow0^+}{\left\{\left.\frac{\left\|\nabla\times\bm{B}_{\bm{f}}-\rho^{-2}\nabla\rho\times\nabla\Pi_{\bm{f}}\right\|}{\left\|\rho^{-2}\nabla\rho\times\nabla\Pi_{\bm{f}}\right\|}\middle/\epsilon\right.\right\}}
	=\lim\limits_{\lambda\rightarrow0^+}\frac{\left\|\nabla\eta\times\left(2\eta\nabla\Pi_{\bm{f}}^{(0)}-\nabla\Pi_{\bm{f}}^{(1)}\right)\right\|}{\left\|\nabla\eta\times\nabla\Pi_{\bm{f}}^{(0)}\right\|},
	\label{eqn:Err_buoy_cent}
\end{align}
which means this buoyancy term is also valid.

\subsubsection{Buoyancy term $-\rho'\bm{u}\cdot\nabla\bm{u}/\rho_0$ corresponding to the convection term}
This buoyancy term is introduced by~\citet{lopez2013boussinesq}. Generally the convection term $\bm{C}=-\bm{u}\cdot\nabla\bm{u}$ is non-conservative:
\begin{align}
	\nabla\Pi_{\bm{C}}^{(0)}\neq\rho_0\bm{C}.
	\label{eqn:Phi_Pi0_conv}
\end{align}
Unfortunately, {\iffalse}(see the supplementary material) {\fi} the relative error of $\bm{B}_{\bm{C}}=-\rho'\bm{u}\cdot\nabla\bm{u}/\rho_0$ is $O(1)$ for general flows:
\begin{align}
	\lim\limits_{\lambda\rightarrow0^+}{\left\{\frac{\left\|\nabla\times\left(\rho'\rho_0^{-1}\bm{C}\right)-\rho^{-2}\nabla\rho\times\nabla\Pi_{\bm{C}}\right\|}{\left\|\rho^{-2}\nabla\rho\times\nabla\Pi_{\bm{C}}\right\|}\right\}}
	=\lim\limits_{\lambda\rightarrow0^+}\frac{\left\|\nabla\times\left[\eta\left(\rho_0\bm{C}-\nabla\Pi_{\bm{C}}^{(0)}\right)\right]\right\|}{\left\|\nabla\eta\times\nabla\Pi_{\bm{C}}^{(0)}\right\|}.
	\label{eqn:Err_buoy_conv}
\end{align}
Therefore for general flows with non-conservative convection, introducing the buoyancy term $\bm{B}_{\bm{C}}=\rho'\bm{C}/\rho_0$ does not help in approximating the original low-Mach-number equations~\eqref{eqn:eqNS}. The main reason is that, the non-conservative part $\bm{C}-\rho_0^{-1}\nabla\Pi_{\bm{C}}$ of the convection term contributes directly (without influencing the baroclinic torque) to $\partial\bm{\omega}/\partial{t}$, and should not appear in the buoyancy term.

\subsubsection{Coriolis buoyancy term $2\rho'\bm{u}\times\bm{\Omega}/\rho_0$}
This buoyancy term is introduced by~\citet{kang2019numerical}. For general three-dimensional (3-D) flows, the Coriolis force $\bm{f}=2\bm{u}\times\bm{\Omega}$ is non-conservative:
\begin{align}
	\nabla\Pi_{\bm{f}}^{(0)}\neq\rho_0\bm{f}.
	\label{eqn:Phi_Pi0_cori}
\end{align}
It could be proved {\iffalse}(see the supplementary material) {\fi}that the relative error of $\bm{B}_{\bm{f}}=2\rho'\bm{u}\times\bm{\Omega}/\rho_0$ does not converge to $0$ even when $\epsilon\rightarrow0$:
\begin{align}
	\lim\limits_{\lambda\rightarrow0^+}{\left\{\frac{\left\|\nabla\times\left(\rho'\rho_0^{-1}\bm{f}\right)-\rho^{-2}\nabla\rho\times\nabla\Pi_{\bm{f}}\right\|}{\left\|\rho^{-2}\nabla\rho\times\nabla\Pi_{\bm{f}}\right\|}\right\}}
	=\lim\limits_{\lambda\rightarrow0^+}\frac{\left\|\nabla\times\left[\eta\left(\rho_0\bm{f}-\nabla\Pi_{\bm{f}}^{(0)}\right)\right]\right\|}{\left\|\nabla\eta\times\nabla\Pi_{\bm{f}}^{(0)}\right\|}.
	\label{eqn:Err_buoy_cori}
\end{align}
Therefore for general 3-D flows with non-conservative Coriolis force, introducing the term $\bm{B}_{\bm{f}}=2\rho'\bm{u}\times\bm{\Omega}/\rho_0$ again does not help in approximating the original low-Mach-number equations~\eqref{eqn:eqNS}. However, for two-dimensional (2-D) flows with very small velocity divergence, the Coriolis force is close to being conservative, and such buoyancy term may be acceptable.

\subsubsection{The $n^\text{th}$-order partial buoyancy term $\bm{B}_{\bm{Y}}^{(n)}$}
For general cases it can be proved {\iffalse}(see the supplementary material) {\fi}that the relative error of $\bm{B}_{\bm{Y}}^{(n)}$ is $O(\epsilon^n)$ convergent:
\begin{align}
	\lim\limits_{\lambda\rightarrow0^+}{\left\{\left.\frac{\left\|\nabla\times\bm{B}_{\bm{Y}}^{(n)}-\rho^{-2}\nabla\rho\times\nabla\Pi_{\bm{Y}}\right\|}{\left\|\rho^{-2}\nabla\rho\times\nabla\Pi_{\bm{Y}}\right\|}\middle/\epsilon^n\right.\right\}}
	=\lim\limits_{\lambda\rightarrow0^+}\frac{\left\|\nabla\eta\times\nabla\Pi_{\bm{Y}}^{(n)}\right\|}{\left\|\nabla\eta\times\nabla\Pi_{\bm{Y}}^{(0)}\right\|}.
	\label{eqn:Err_buoy_new}
\end{align}
Therefore the $n^\text{th}$-order partial buoyancy term with small $\epsilon$ could be made arbitrarily accurate by increasing $n$. However, this generally requires solving more Poisson equations.

\subsubsection{The $n^\text{th}$-order Type-I buoyancy term $\bm{B}^{(n)}$}
It can be proved {\iffalse}(see the supplementary material) {\fi}that for general cases, the relative error of $\bm{B}^{(n)}$ is $O(\epsilon^n)$ convergent:
\begin{align}
	\lim\limits_{\lambda\rightarrow0^+}{\left\{\left.\frac{\left\|\nabla\times\bm{B}^{(n)}-\rho^{-2}\nabla\rho\times\nabla\Pi\right\|}{\left\|\rho^{-2}\nabla\rho\times\nabla\Pi\right\|}\middle/\epsilon^n\right.\right\}}
	=\lim\limits_{\lambda\rightarrow0^+}\frac{\left\|\nabla\eta\times\nabla\Pi^{(n)}\right\|}{\left\|\nabla\eta\times\nabla\Pi^{(0)}\right\|}.
	\label{eqn:Err_buoy_T1}
\end{align}

\subsubsection{The $n^\text{th}$-order Type-II buoyancy term $\hat{\bm{B}}^{(n)}$}

Consider a set of buoyancy-driven flows with $\bm{f}=\nabla\Phi$ and $\left(\|\bm{f}\|/\|\bm{A}+\bm{C}+\bm{D}\|\right)\sim O(\epsilon^{-1})$. It can be proved {\iffalse}(see the supplementary material) {\fi}that in this circumstance the relative error of $\hat{\bm{B}}^{(n)}$ is generally $O(\epsilon^n)$ convergent:
\begin{equation}
	\begin{aligned}
		&\lim\limits_{\lambda\rightarrow0^+}{\left\{\left.\frac{\left\|\nabla\times\hat{\bm{B}}^{(n)}-\rho^{-2}\nabla\rho\times\nabla\Pi\right\|}{\left\|\rho^{-2}\nabla\rho\times\nabla\Pi\right\|}\middle/\epsilon^n\right.\right\}}
		\\
		=&\lim\limits_{\lambda\rightarrow0^+}\frac{\left\|\nabla\eta\times\left[\nabla\Pi_{\bm{f}}^{(n)}+\epsilon^{-1}\left(\nabla\Pi_{\bm{A}}^{(n-1)}+\nabla\Pi_{\bm{C}}^{(n-1)}+\nabla\Pi_{\bm{D}}^{(n-1)}\right)\right]\right\|}{\left\|\nabla\eta\times\nabla\Pi_{\bm{f}}^{(0)}\right\|}.
	\end{aligned}
	\label{eqn:Err_buoy_T2}
\end{equation}
This indicates that for such buoyancy-driven flows, $\hat{\bm{B}}^{(n)}$ has the same scaling of the relative error as $\bm{B}^{(n)}$, although it requires solving less Poisson equations. Therefore, although we cannot avoid solving extra equations for accurate approximation of the baroclinic torque, computational costs could be properly reduced based on the amplitudes of $\{\bm{A},\bm{C},\bm{D},\bm{f}\}$.

\subsection{Invariance of buoyancy terms under frame transformations}\label{sec:analytic_invariance}

Since body forces and some other terms in the momentum equation may be different in different translating or rotating frames, the invariance of buoyancy terms in different inertial and non-inertial frames should be examined. {\iffalse}Results are shown here and the detailed derivations are shown in the supplementary material.{\fi}

\subsubsection{Static frame and translating frame}\label{sec:analytic_invariance_translate}

Consider a translating frame moving with speed $\bm{v}_0+\bm{a}(t-t_0)$ relative to the static frame. $\bm{v}_0$ and $\bm{a}$ are assumed to be constant. The flow variables in the static inertial frame are marked with a subscript ``I'', while those in the translating frame have no subscript. Assuming that at $t=t_0$ the static frame and the translating frame coincide, the variables have the following transformation
\begin{align}
	\begin{aligned}
		\bm{u}_I(\bm{x},t)&=\bm{u}\left(\bm{x}-\bm{v}_0(t-t_0)-\bm{a}(t-t_0)^2/2,t\right)+\bm{v}_0+\bm{a}(t-t_0),
		\\
		\rho_I(\bm{x},t)&=\rho\left(\bm{x}-\bm{v}_0(t-t_0)-\bm{a}(t-t_0)^2/2,t\right).
	\end{aligned}
\label{eqn:urho_transf_translate}
\end{align}
Apparently, in the translating frame there is a non-inertial force $\bm{f}=-\bm{a}$ similar as a gravity force, and it could be used to construct a buoyancy term $-\rho'\bm{a}/\rho_0$ in the translating frame following~\citet{boussinesq1903theorie}. However, $-\rho'\bm{a}/\rho_0$ could not be found in the static frame where the body force $\bm{f}_I=0$. The buoyancy term corresponding to the convection term~\citep{lopez2013boussinesq} is also not invariant under the present transformation:
\begin{align}
	\bm{C}_I=\bm{C}-\bm{v}_0\cdot\nabla\bm{u}.
	\label{eqn:conv_transf_translate}
\end{align}

For an $n^{\text{th}}$-order type-I buoyancy term $\bm{B}^{(n)}$ ($n\geq1$, see equation~\eqref{eqn:T1buoy} in \S\ref{sec:analytic_perturbation}), it can be proved that
\begin{align}
	\bm{B}_I^{(n)}=\bm{B}^{(n)},
	\label{eqn:Bn_ACDf_invariant_translate}
\end{align}
which means that the $n^{\text{th}}$-order type-I buoayncy term is invariant under the present transformation.

\subsubsection{Static frame and rotating frame}

For simplicity, assume that the angular velocity $\bm{\Omega}$ of the rotating frame is constant, $z$ axis is the rotating axis, and the two frames coincide at $t=t_0$. Again, the flow variables in the static inertial frame are marked with a subscript ``I'', while those in the rotating frame have no subscript. The variables between two frames have the following transformation:
\begin{align}
	u_{r,I}\left(r,\phi,z,t\right)&=u_{r}\left(r,\phi-\Omega(t-t_0),z,t\right),
	\notag \\
	u_{\phi,I}\left(r,\phi,z,t\right)&=u_{\phi}\left(r,\phi-\Omega(t-t_0),z,t\right)+\Omega r,
	\notag \\
	u_{z,I}\left(r,\phi,z,t\right)&=u_{z}\left(r,\phi-\Omega(t-t_0),z,t\right),
	\label{eqn:urho_transf_rotate} \\
	\rho_I\left(r,\phi,z,t\right)&=\rho\left(r,\phi-\Omega(t-t_0),z,t\right).
	\notag
\end{align}

It can be proved that, the buoyancy term $\rho'\bm{C}_I/\rho_0$ constructed with the convection term in the static frame~\citep{lopez2013boussinesq}, is not equal to the summation of the buoyancy terms corresponding to the convection term, centrifugal force and Coriolis force in the rotating frame:
\begin{align}
	\bm{C}_I
	=\bm{C}+\Omega^2r\bm{e}_r+2\Omega\bm{u}\times\bm{e}_z-\Omega\left(\frac{\partial{u_r}}{\partial\phi}\bm{e}_r+\frac{\partial{u_{\phi}}}{\partial\phi}\bm{e}_{\phi}+\frac{\partial{u_z}}{\partial\phi}\bm{e}_z\right).
	\label{eqn:conv_transf_rotate}
\end{align}

For an $n^{\text{th}}$-order type-I buoyancy term $\bm{B}^{(n)}$ ($n\geq1$), the invariance under this frame transformation could be proved:
\begin{align}
	\bm{B}_I^{(n)}=\bm{B}^{(n)},
	\label{eqn:Bn_ACDf_invariant_rotate}
\end{align}
similarly as the procedure corresponding to the frame transformation in~\S\ref{sec:analytic_invariance_translate}.
\\\ \\

For the type-II buoyancy terms, although they are sometimes as accurate as the type-I terms with the same orders, they might not have the invariance under frame transformations. Therefore, none of the previous and newly introduced buoyancy terms are invariant under the two frame transformations mentioned above, except for the type-I buoyancy terms.

\section{Numerical validation}\label{sec:numerical}
Here we will use the numerical simulation of a vertical convection~{\citep{ng2015vertical,shishkina2016momentum}} for an \textit{a posteriori} validation of the derivations in \S\ref{sec:analytic}. Since we are only considering about the error, both type-I and type-II buoyancy terms can be used, and we will focus on the type-II buoyancy terms due to their higher computational efficiency.  It is expected that the simulation results of the modified equations~\eqref{eqn:eqNS_buoy} equipped with the type-II buoyancy terms~\eqref{eqn:T2buoy} could converge to the simulation results of the original low-Mach-number equations~\eqref{eqn:eqNS} with relative errors of $O(\epsilon^n)$.

\subsection{Original equations}\label{sec:numerical_original}
\begin{figure}
	\centering
	\includegraphics[width=0.5\textwidth]{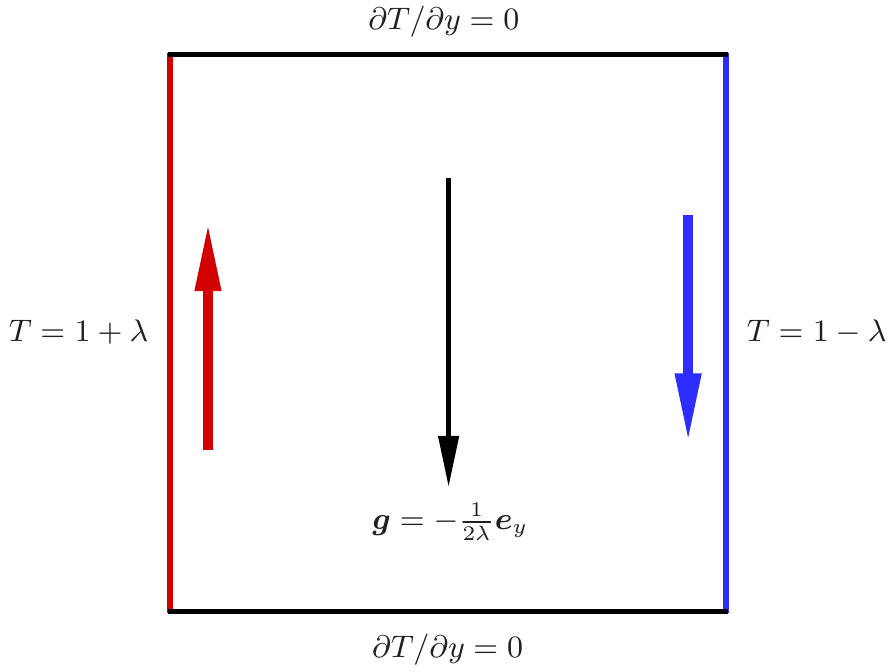}
	\caption{Sketch of the vertical convection in 2-D square cavity.}
	\label{fig:Sketch}
\end{figure}

The flow set-up and original governing equations are completely equivalent to those in~\citet{wang2019non}. Consider a 2-D square cavity with height $H^*$ and width $L^*=H^*$. $x^*,y^*$ are coordinates of horizontal and vertical directions respectively, and $u^*,v^*$ are velocity components in $x^*,y^*$ directions respectively. $T^*$ is the temperature, with left and right walls having constant temperatures $T^*_h$ and $T^*_c$ respectively, and horizontal walls are adiabatic. Assume that $T^*_h>T^*_c$ and define the reference temperature $T^*_r=(T^*_h+T^*_c)/2=300\text{K}$, so that the parameter $\lambda$ could be defined as $(T^*_h-T^*_r)/T^*_r$ to represent the scale of relative temperature variation. In addition, we define gravity acceleration $\bm{g}^*=-g^*\bm{e}_y$, reference density $\rho^*_r$, reference dynamic viscosity $\mu_r^*$, reference thermal conductivity $\kappa_r^*$, reference isobaric specific heat $c_{p,r}^*$, universal gas constant $R^*=8.314\text{JK}^{-1}\text{mol}^{-1}$, reference velocity $u_r^*=(2\lambda g^*L^*)^{1/2}$, reference time scale $t_r^*=L^*/u_r^*$, and reference thermodynamic pressure $p_r^*=\rho_r^*R^*T_r^*$. Following the variable non-dimensionalization:
\begin{gather}
	\bm{x}=\bm{x}^*/L^*,\ \bm{u}=\bm{u}^*/u_r^*,\ t=t^*/t_r^*,
	\notag \\
	\rho=\rho^*/\rho_r^*,\ p=p^*/p_r^*,\ \Pi=\Pi^*/(\rho_r^*u_r^{*2}),\ T=T^*/T_r^*,
	\label{eqn:variable_nondimen} \\
	\mu=\mu^*/\mu_r^*,\ \kappa=\kappa^*/\kappa_r^*,\ c_p=c_p^*/c_{p,r}^*,\ \bm{g}=\bm{g}^*/(L^{*-1}u_r^{*2})=-(2\lambda)^{-1}\bm{e}_y,
	\notag
\end{gather}
the non-dimensionalized low-Mach-number equations and boundary conditions could be written as~\citep{wang2019non}:
\begin{subequations}
\begin{gather}
	\frac{1}{\rho}\frac{d\rho}{dt}+\nabla\cdot\bm{u}=0,
	\\
	\frac{\partial\bm{u}}{\partial{t}}+\bm{u}\cdot\nabla\bm{u}=-\frac{1}{\rho}\nabla\Pi+\frac{1}{\rho}\nabla\cdot\mathbb{T}+\bm{g},
	\\
	\frac{\partial{T}}{\partial{t}}+\bm{u}\cdot\nabla{T}=-\frac{1}{\rho c_p}\nabla\cdot\bm{J}+\frac{\Gamma}{\rho c_p}\frac{dp}{dt},
	\\
	p=\rho T,
	\\
	x=\pm0.5:\ \bm{u}=0,\ T=1\mp\lambda,
	\\
	y=\pm0.5:\ \bm{u}=0,\ \partial{T}/\partial{y}=0.
\end{gather}
\label{eqn:eqNS_thermal}
\end{subequations}
A sketch illustrating the boundary conditions is shown in Figure~\ref{fig:Sketch}.
The viscous stress tensor and heat flux are
\begin{align}\begin{aligned}
	\mathbb{T}=&\mu\left[\nabla\bm{u}+\left(\nabla\bm{u}\right)^T\right]-\frac{2\mu}{3}\mathbb{I}\nabla\cdot\bm{u},
	\\
	\bm{J}=&-\kappa\nabla T
	\label{eqn:Tau_J},
\end{aligned}\end{align}
respectively, where $\mu,\ \kappa$ follow the Sutherland laws
\begin{align}\begin{aligned}
		\mu=&\sqrt{\frac{Pr}{Ra}}T^{1.5}\frac{1+S_{\mu}}{T+S_{\mu}},
		\\
		\kappa=&\sqrt{\frac{1}{RaPr}}T^{1.5}\frac{1+S_{\kappa}}{T+S_{\kappa}},
		\label{eqn:sutherland}
\end{aligned}\end{align}
with the Rayleigh number $Ra$ and Prandtl number $Pr$ defined as
\begin{align}
	Ra=\frac{2\lambda c_{p,r}^*\rho_r^{*2}g^*L^{*3}}{\mu_r^*\kappa_r^*}
	,\
	Pr=\frac{c_{p,r}^*\mu_r^*}{\kappa_r^*},
	\label{eqn:RaPr}
\end{align}
and with $S_{\mu}=0.368,\ S_{\kappa}=0.648$ corresponding to air at $T_r^*=300\text{K}$. The isobaric specific heat $c_p$ is assumed to be $1$, and the thermodynamic pressure $p(t)$ is assumed to be uniform in space~\citep{wang2019non}. Noticing that the total mass $M=\int_{\mathcal{V}}{\rho dV}$ is constant in $\mathcal{V}=[-0.5,0.5]\times[-0.5,0.5]$, the thermodynamic pressure could be computed using the temperature field:
\begin{align}
	p=M\left(\int_{\mathcal{V}}{\frac{dV}{T}}\right)^{-1}.
	\label{eqn:thermo_pressure}
\end{align}
For the thermal resilience term $\Gamma dp/dt$, $\gamma=1.4$ is the specific heat ratio and $\Gamma=(\gamma-1)/\gamma$ is a measure of the resilience of the fluid. Using equation (\ref{eqn:eqNS_thermal}a), (\ref{eqn:eqNS_thermal}c), (\ref{eqn:eqNS_thermal}d), \eqref{eqn:thermo_pressure} and the non-penetrative boundary condition, the time derivative of the thermodynamic pressure could be derived:
\begin{align}
	\frac{dp}{dt}=-\frac{1}{(1-\Gamma)V}\int_{\mathcal{V}}{\nabla\cdot\bm{J}dV}.
	\label{eqn:dpdt}
\end{align}
Therefore the source term of density (R.H.S. term of \eqref{eqn:eqdensity}) does not explicitly contain velocity:
\begin{align}
	\frac{d\rho}{dt}=Q=\frac{1}{T}\left(\nabla\cdot\bm{J}-\frac{1}{V}\int_{\mathcal{V}}{\nabla\cdot\bm{J}dV}\right).
	\label{eqn:Q_no_u}
\end{align}

Equations~\eqref{eqn:eqNS_thermal} are solved using a second-order central difference code on a uniform grid. The corresponding hydrodynamic pressure equation~\eqref{eqn:Pi} is solved using the successive over-relaxation (SOR) method. Using a time-stepping approach, steady solutions $\left(\bar{\bm{u}}(\bm{x}),\bar{T}(\bm{x})\right)$ are achieved in the sense of machine precision. The code is validated at $Ra=10^6$, $Pr=0.71$ and varying $\lambda$. The Oberbeck-Boussinesq (OB) approximation could be simulated in the present code with $\lambda\approx0$. With increasing $\lambda$, the non-Oberbeck-Boussinesq effect will come out and it is significant in the cases with $\lambda=0.2, 0.4, 0.6$. The Nusselt numbers $Nu$ from the present simulations are compared with the reference values from~\citet{wang2019non}. As shown in Table~\ref{tab:validation}, the present code can accurately predict $Nu$ for all cases with relative deviations less than $0.1\%$, demonstrating the correctness of the present code in a wide range of $\lambda$.

Using the present code corresponding to equation~\eqref{eqn:eqNS_thermal}, thirteen simulations are performed on a uniform $256\times256$ grid at $Ra=10^6,\ Pr=0.71$ and $\lambda=0.6\times(2/3)^n$ with integer $n$ ranging from $0$ to $12$. The $13$ cases are regarded as the reference cases with the steady flow fields $\left(\bar{\bm{u}}(\bm{x};\lambda),\bar{T}(\bm{x};\lambda)\right)$, which will be used to assess the results corresponding to different orders of the type-II buoyancy terms.

\begin{table}
		\begin{center}
			\begin{tabular}{m{2.7cm}<{\centering} m{2.2cm}<{\centering} m{2.2cm}<{\centering} m{2.2cm}<{\centering} m{2.2cm}<{\centering}}
				\toprule
				$Ra$ & $10^6$(OB) & $10^6$($\lambda=0.2$) & $10^6$($\lambda=0.4$) & $10^6$($\lambda=0.6$) \\
				\midrule
				$Nu$ &  $8.832$ & $8.814$ & $8.735$ & $8.597$ \\
				$Nu_{\text{ref}}$ &  $8.830$ & $8.806$ & $8.732$ & $8.602$ \\
				\midrule
                $|Nu-Nu_{\text{ref}}|/Nu_{\text{ref}}$  & $2.3\times10^{-4}$ & $9.1\times10^{-4}$ & $3.4\times10^{-4}$ & $5.8\times10^{-4}$\\
				\bottomrule
			\end{tabular}
		\end{center}
		\caption{Comparison of the Nusselt numbers $Nu$ at $Ra=10^6$, $Pr=0.71$ and varying $\lambda$ from present simulations and the reference works from \citet{wang2019non}. The grid is uniform and its number is $256\times256$.}
		\label{tab:validation}
\end{table}


\subsection{Modified equations}\label{sec:numerical_modified}

In order to apply the type-II buoyancy terms, the modified equations can be obtained following the procedure described in \S\ref{sec:analytic_buoyancy}:
\begin{subequations}
	\begin{gather}
		\frac{1}{\rho}\frac{d\rho}{dt}+\nabla\cdot\bm{u}=0,
		\\
		\frac{\partial\bm{u}}{\partial{t}}+\bm{u}\cdot\nabla\bm{u}=-\frac{1}{\rho_0}\nabla\widetilde{\Pi}+\frac{1}{\rho}\nabla\cdot\mathbb{T}+\bm{g}+\bm{B},
		\\
		\frac{\partial{T}}{\partial{t}}+\bm{u}\cdot\nabla{T}=-\frac{1}{\rho c_p}\nabla\cdot\bm{J}+\frac{\Gamma}{\rho c_p}\frac{dp}{dt},
		\\
		p=\rho T,
		\\
		x=\pm0.5:\ \bm{u}=0,\ T=1\mp\lambda,
		\\
		y=\pm0.5:\ \bm{u}=0,\ \partial{T}/\partial{y}=0,
	\end{gather}
	\label{eqn:eqNS_thermal_buoy}
\end{subequations}
with the basic parameters and equations~\eqref{eqn:Tau_J}-\eqref{eqn:Q_no_u} being the same as \S\ref{sec:numerical_original}. $\rho_0$ is chosen as $p/T_r$, so that $\epsilon=\max_{\mathcal{V}}\left[|\rho'/\rho|\right]\equiv\lambda$. The definitions of $\bm{A},\bm{C},\bm{D},\bm{f}$ and the corresponding hydrodynamic pressures are the same as those defined in \S\ref{sec:analytic_LMN}.

Since $\|\bm{f}\|\sim O(\epsilon^{-1})$ and $\|\bm{A}+\bm{C}+\bm{D}\|\sim O(1)$, an $n^{\text{th}}$-order type-II buoyancy term $\hat{\bm{B}}^{(n)}$ has a relative error of $O(\epsilon^n)$ according to \S\ref{sec:analytic_accuracy}. Apparently the error of $\hat{\bm{B}}^{(n)}$ is also $O(\epsilon^n)$. Four groups of steady flow fields are defined corresponding to different orders of type-II buoyancy terms:
\begin{subequations}
	\begin{align}
 \left(\bar{\bm{u}}^{(1)},\bar{T}^{(1)}\right):&\
    \hat{\bm{B}}^{(1)}=\frac{\rho'}{\rho}\bm{g},
	 \label{eqn:u1_buoy}\\
 \left(\bar{\bm{u}}^{(2)},\bar{T}^{(2)}\right):&\
	\hat{\bm{B}}^{(2)},
	\label{eqn:u2_buoy} \\
 \left(\bar{\bm{u}}^{(3)},\bar{T}^{(3)}\right):&\
	\hat{\bm{B}}^{(3)},
	\label{eqn:u3_buoy} \\
 \left(\bar{\bm{u}}^{(\infty)},\bar{T}^{(\infty)}\right):&\	\hat{\bm{B}}^{(\infty)}=\bm{B}^{\dagger}.
	\label{eqn:u4_buoy}
	\end{align}
\label{eqn:u1234_buoy}
\end{subequations}
It can be seen that $\hat{\bm{B}}^{(1)}$ is valid with $O(\epsilon)$ error, and does not require solving any extra Poisson equation. $\hat{\bm{B}}^{(2)}$ is more accurate with $O(\epsilon^2)$ error, but requires solving one extra Poisson equation for each time step. Similarly, $\hat{\bm{B}}^{(3)}$ has $O(\epsilon^3)$ error but requires solving two extra Poisson equations for each time step. $\hat{\bm{B}}^{(\infty)}$ theoretically requires solving infinitely many times of Poisson equations for each time step.

Equations~\eqref{eqn:eqNS_thermal_buoy} are solved using a finite-difference code which is almost the same as that described in \S\ref{sec:numerical_original}, except that the Poisson equations are decoupled using discrete cosine transform (DCT) and solved directly~\citep{zhang2015direct,zhang2020controlling}. The simulation results using this code and the Boussinesq approximation are compared with those in~\citet{de1983natural} or~\citet{wang2019non} for $Ra=10^4, 10^5, 10^6$ and $Pr=0.71$, and the relatives errors are again less than $0.1\%$. Although an infinite sequence of Poisson equations are required for computing $\hat{\bm{B}}^{(\infty)}$ theoretically, solving $100$ extra Poisson equations for each time step is sufficient to make the residual converge to $0$ in the sense of machine precision for $\epsilon\leq0.6$. Each group of flow fields described by equations~\eqref{eqn:u1_buoy}-\eqref{eqn:u4_buoy} corresponds to $13$ cases described in \S\ref{sec:numerical_original}, that is, $Ra=10^6$, $Pr=0.71$ and $\lambda=0.6\times(2/3)^n$ with integer $n$ ranging from $0$ to $12$.

\subsection{Simulation results}\label{sec:numerical_result}
\begin{figure}
	\centering
	\includegraphics[width=0.9\textwidth]{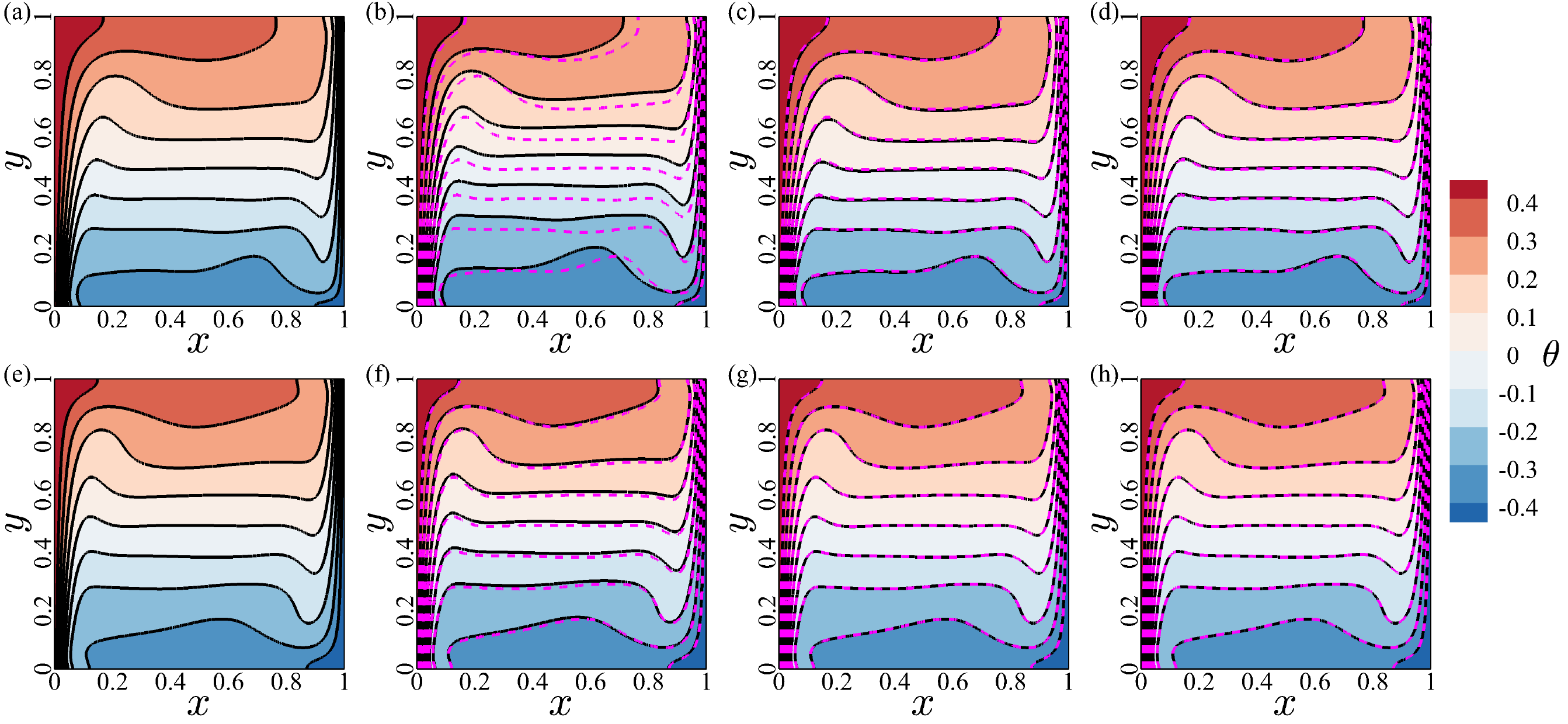}
	\caption{Contour of normalized temperature $\theta$. Isolines correspond to values $-0.4:(0.1):0.4$. (a-d) $\lambda=0.6$. (e-h) $\lambda=0.178$. (a,e) Contour and solid isolines of $\bar{\theta}$. (b,f) Contour and solid isolines of $\bar{\theta}^{(1)}$. (c,g) Contour and solid isolines of $\bar{\theta}^{(2)}$. (d,h) Contour and solid isolines of $\bar{\theta}^{(3)}$. (b-d,f-h) Purple dashed isolines of $\bar{\theta}$ of the corresponding $\lambda$. }
	\label{fig:T_compare}
\end{figure}

\begin{figure}
	\centering
	\includegraphics[width=0.9\textwidth]{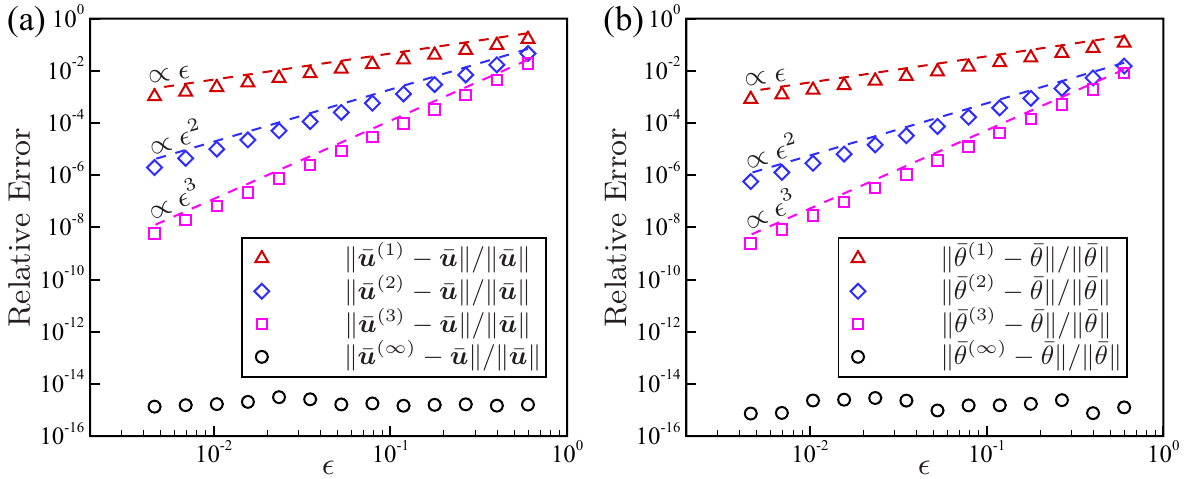}
	\caption{Relative errors of steady flow fields. (a) Relative errors of the velocity fields. (b) Relative errors of the temperature fields.}
	\label{fig:Error}
\end{figure}

A normalized temperature is defined as $\theta=(T-1)/2\lambda$. Figure~\ref{fig:T_compare} shows the contours of $\bar{\theta},\bar{\theta}^{(1)},\bar{\theta}^{(2)},\bar{\theta}^{(3)}$, with parameter $\lambda=0.6$ and $\lambda=0.178$. Here $\lambda=0.6$ corresponds to a large density variation ($\max_{\mathcal{V}}[\rho]/\min_{\mathcal{V}}[\rho]\equiv4$), and $\lambda=0.178$ corresponds to a medium density variation ($\max_{\mathcal{V}}[\rho]/\min_{\mathcal{V}}[\rho]\equiv1.43$). For a large density variation ($\lambda=0.6$), Figure~\ref{fig:T_compare}(b) indicates that, the first-order type-II buoyancy term $\hat{\bm{B}}^{(1)}$ is a very coarse approximation and could only predict qualitative properties of the steady flow. To improve the accuracy, $\hat{\bm{B}}^{(2)}$ with a higher order could be used, and Figure~\ref{fig:T_compare}(c) shows that the error is very small. It should be emphasized that $\hat{\bm{B}}^{(2)}$ only requires solving one extra Poisson equation. For even better approximation, $\bm{B}^{(3)}$ could be used (Figure~\ref{fig:T_compare}(d)) but it requires solving two extra Poisson equations. Fortunately, when the density variation is not so large ($\lambda=0.178$), solving one extra Poisson equations seems sufficient, because the error of $\hat{\bm{B}}^{(2)}$ is already unnoticeable, as indicated by Figure~\ref{fig:T_compare}(g).

To show the accuracy of type-II buoyancy terms and the ``exact'' buoyancy term in detail, the relative errors of steady $\bar{\bm{u}}$ and $\bar{\theta}$ fields corresponding to buoyancy terms $\hat{\bm{B}}^{(1)},\hat{\bm{B}}^{(2)},\hat{\bm{B}}^{(3)},\hat{\bm{B}}^{(\infty)}$ are shown in Figure~\ref{fig:Error}. It is clearly seen that using the type-II buoyancy terms $\hat{\bm{B}}^{(1)},\hat{\bm{B}}^{(2)},\hat{\bm{B}}^{(3)}$ with errors $O(\epsilon)$, $O(\epsilon^2)$ and $O(\epsilon^3)$ respectively, the relative errors of $(\bar{\bm{u}},\bar{\theta})$ are $O(\epsilon)$, $O(\epsilon^2)$ and $O(\epsilon^3)$ correspondingly. Since $\|\bar{\bm{u}}\|$ is $O(1)$, we may arrive at the conclusion that the \textit{a posteriori} errors of steady velocity fields have the same scaling as the theoretical errors of the corresponding buoyancy terms. For the ``exact'' buoyancy term $\bm{B}^{\dagger}=\hat{\bm{B}}^{(\infty)}$ with no error theoretically, it can also be seen from Figure~\ref{fig:Error} that the steady flow fields $(\bar{\bm{u}}^{(\infty)},\bar{T}^{(\infty)})$ are equal to $(\bar{\bm{u}},\bar{T})$ in the sense of machine precision. From the above results, we have confirmed by numerical simulations that for $\epsilon\leq0.6$ ($\max_{\mathcal{V}}[\rho]/\min_{\mathcal{V}}[\rho]\leq4$), the modified equation~\eqref{eqn:eqNS_thermal_buoy} with the ``exact'' buoyancy term $\bm{B}^{\dagger}=\hat{\bm{B}}^{(\infty)}$, is equivalent to the original low-Mach-number equations~\eqref{eqn:eqNS_thermal}.

Figure~\ref{fig:Error} can further suggest the valid $\epsilon$ range for each buoyancy term to achieve the required accuracy. For example, if the error bound is $1\%$ for both the velocity and temperature fields, $\hat{\bm{B}}^{(1)}$ could only be used for $\epsilon\leq0.035$, while the more accurate $\hat{\bm{B}}^{(2)}$ could be used for $\epsilon\leq0.267$, and $\hat{\bm{B}}^{(3)}$ is valid for $\epsilon\leq0.4$.

It should be noted that the spectral Poisson solver could be much more efficient than many iterative solvers for the present problem size, although the type-II buoyancy terms often require solving extra Poisson equations. Therefore, the simulation using the modified equations with type-II buoyancy terms can be sometimes more efficient than that using the original low-Mach-number equations.

\section{Conclusion}\label{sec:conclusion}

In the present paper, a regular perturbation method is applied to the hydrodynamic pressure equation, leading to a sequence of Poisson equations. The solutions of the Poisson equations form a series expansion of baroclinic torque and thus the error analysis of any buoyancy term can be performed, which could help to recognize and prevent invalid buoyancy terms. In addition, new types of buoyancy terms, namely the $n^{\text{th}}$-order partial buoyancy terms, type-I buoyancy terms, and type-II buoyancy terms, are then constructed with the perturbation solutions, and they are proved to be accurate theoretically and numerically.

Although the low-Mach-number equations can be used in the numerical simulations of flows with significant density variation, they may cause troubles in theoretical analysis due to the coupling of the pressure and density. The present perturbation analysis of the baroclinic torque allows to modify the momentum equation of the original low-Mach-number equations by introducing new types of buoyancy terms, and the hydrodynamic pressure in the modified momentum equation satisfies the Poisson equation. This could greatly simplify some theoretical analysis because it only requires solving the Poisson equations for the momentum equation and the buoyancy term. In addition, with new types of buoyancy terms, some numerical codes with spectral Poisson solvers can be used to simulate flows with significant density variation.

\section*{Acknowledgement}
This work was supported by the National Science Foundation of China (NSFC grant nos 11822208, 11988102, 11772297 and 91852205). The authors would like to thank Professor D. Lohse for many useful suggestions.

\section*{Authors conflict}
The authors report no conflict of interest.

\bibliographystyle{jfm}
\bibliography{Reference}

\begin{thebibliography}{29}
\expandafter\ifx\csname natexlab\endcsname\relax\def\natexlab#1{#1}\fi
\def\au#1{#1} \def\ed#1{#1} \def\yr#1{#1}\def\at#1{#1}\def\jt#1{\textit{#1}}
  \def\bt#1{#1}\def\bvol#1{\textbf{#1}} \def\vol#1{#1} \def\pg#1{#1}
  \def\publ#1{#1}\def\arxiv#1{#1}\def\org#1{#1}\def\st#1{\textit{#1}}

\bibitem[Ahlers {\em et~al.\/}(2009)Ahlers, Grossmann \& Lohse]{ahlers2009heat}
{\sc \au{Ahlers, G.}, \au{Grossmann, S.} \& \au{Lohse, D.}} \yr{2009}  \at{Heat
  transfer and large scale dynamics in turbulent {Rayleigh-B{\'e}nard}
  convection}.  \jt{Rev.~Mod.~Phys.}  \bvol{81}~(2),  \pg{503--537}.

\bibitem[Barcilon \& Pedlosky(1967)]{barcilon1967steady}
{\sc \au{Barcilon, V.} \& \au{Pedlosky, J.}} \yr{1967}  \at{On the steady
  motions produced by a stable stratification in a rapidly rotating fluid}.
  \jt{J.~Fluid Mech.}  \bvol{29}~(4),  \pg{673--690}.

\bibitem[Boussinesq(1903)]{boussinesq1903theorie}
{\sc \au{Boussinesq, J.}} \yr{1903} {\em Theorie Analytique de la Chaleur vol
  2\/}.  \publ{Paris: Gauthier-Villars}.

\bibitem[Busse \& Carrigan(1974)]{busse1974convection}
{\sc \au{Busse, F.~H.} \& \au{Carrigan, C.~R.}} \yr{1974}  \at{Convection
  induced by centrifugal buoyancy}.  \jt{J.~Fluid Mech.}  \bvol{62}~(3),
  \pg{579--592}.

\bibitem[Chandrasekhar(2013)]{chandrasekhar2013hydrodynamic}
{\sc \au{Chandrasekhar, S.}} \yr{2013} {\em Hydrodynamic and hydromagnetic
  stability\/}.  \publ{Courier Corporation}.

\bibitem[Gray \& Giorgini(1976)]{gray1976validity}
{\sc \au{Gray, D.~D.} \& \au{Giorgini, A.}} \yr{1976}  \at{The validity of the
  boussinesq approximation for liquids and gases}.
  \jt{Int.~J.~Heat~Mass~Trans.}  \bvol{19}~(5),  \pg{545--551}.

\bibitem[Homsy \& Hudson(1969)]{homsy1969centrifugally}
{\sc \au{Homsy, G.~M.} \& \au{Hudson, J.~L.}} \yr{1969}  \at{Centrifugally
  driven thermal convection in a rotating cylinder}.  \jt{J.~Fluid Mech.}
  \bvol{35}~(1),  \pg{33--52}.

\bibitem[Horn \& Aurnou(2018)]{horn2018regimes}
{\sc \au{Horn, S.} \& \au{Aurnou, J.~M.}} \yr{2018}  \at{Regimes of
  coriolis-centrifugal convection}.  \jt{Phys.~Rev.~Lett.}  \bvol{120}~(20),
  \pg{204502}.

\bibitem[Jahanbakhshi {\em et~al.\/}(2015)Jahanbakhshi, Vaghefi \&
  Madnia]{jahanbakhshi2015baroclinic}
{\sc \au{Jahanbakhshi, R.}, \au{Vaghefi, N.~S.} \& \au{Madnia, C.~K.}}
  \yr{2015}  \at{Baroclinic vorticity generation near the turbulent/
  non-turbulent interface in a compressible shear layer}.  \jt{Phys.~Fluids}
  \bvol{27}~(10),  \pg{105105}.

\bibitem[Kang {\em et~al.\/}(2019)Kang, Meyer, Yoshikawa \&
  Mutabazi]{kang2019numerical}
{\sc \au{Kang, C.}, \au{Meyer, A.}, \au{Yoshikawa, H.~N.} \& \au{Mutabazi, I.}}
  \yr{2019}  \at{Numerical study of thermal convection induced by centrifugal
  buoyancy in a rotating cylindrical annulus}.  \jt{Phys.~Rev.~Fluids}
  \bvol{4}~(4),  \pg{043501}.

\bibitem[Kang {\em et~al.\/}(2015)Kang, Yang \& Mutabazi]{kang2015thermal}
{\sc \au{Kang, C.}, \au{Yang, K.-S.} \& \au{Mutabazi, I.}} \yr{2015}
  \at{Thermal effect on large-aspect-ratio couette--taylor system: numerical
  simulations}.  \jt{J.~Fluid Mech.}  \bvol{771},  \pg{57--78}.

\bibitem[Livescu(2020)]{livescu2020turbulence}
{\sc \au{Livescu, D.}} \yr{2020}  \at{Turbulence with large thermal and
  compositional density variations}.  \jt{Annu.~Rev.~Fluid~Mech.}  \bvol{52}.

\bibitem[Livescu \& Ristorcelli(2007)]{livescu2007buoyancy}
{\sc \au{Livescu, D.} \& \au{Ristorcelli, J.~R.}} \yr{2007}
  \at{Buoyancy-driven variable-density turbulence}.  \jt{J.~Fluid Mech.}
  \bvol{591},  \pg{43--71}.

\bibitem[Lohse \& Xia(2010)]{lohse2010small}
{\sc \au{Lohse, D.} \& \au{Xia, K.-Q.}} \yr{2010}  \at{Small-scale properties
  of turbulent {Rayleigh-B{\'e}nard} convection}.  \jt{Annu.~Rev.~Fluid~Mech.}
  \bvol{42},  \pg{335--364}.

\bibitem[Lopez {\em et~al.\/}(2013)Lopez, Marques \&
  Avila]{lopez2013boussinesq}
{\sc \au{Lopez, J.~M.}, \au{Marques, F.} \& \au{Avila, M.}} \yr{2013}  \at{The
  boussinesq approximation in rapidly rotating flows}.  \jt{J.~Fluid Mech.}
  \bvol{737},  \pg{56--77}.

\bibitem[Majda \& Sethian(1985)]{majda1985derivation}
{\sc \au{Majda, A.} \& \au{Sethian, J.}} \yr{1985}  \at{The derivation and
  numerical solution of the equations for zero mach number combustion}.
  \jt{Combust. Sci. Tech.}  \bvol{42}~(3-4),  \pg{185--205}.

\bibitem[Mcmurtry {\em et~al.\/}(1989)Mcmurtry, Riley \&
  Metcalfe]{mcmurtry1989effects}
{\sc \au{Mcmurtry, P.~A.}, \au{Riley, J.~J.} \& \au{Metcalfe, R.~W.}} \yr{1989}
   \at{Effects of heat release on the large-scale structure in turbulent mixing
  layers}.  \jt{J.~Fluid Mech.}  \bvol{199},  \pg{297--332}.

\bibitem[Ng {\em et~al.\/}(2015)Ng, Ooi, Lohse \& Chung]{ng2015vertical}
{\sc \au{Ng, C.~S.}, \au{Ooi, A.}, \au{Lohse, D.} \& \au{Chung, D.}} \yr{2015}
  \at{Vertical natural convection: application of the unifying theory of
  thermal convection}.  \jt{J.~Fluid Mech.}  \bvol{764},  \pg{349--361}.

\bibitem[Paolucci(1982)]{paolucci1982filtering}
{\sc \au{Paolucci, S.}} \yr{1982} {\em Filtering of sound from the
  Navier-Stokes equations\/}.  \publ{Sandia National Laboratories Livermore,
  CA, USA}.

\bibitem[Paolucci(1990)]{paolucci1990direct}
{\sc \au{Paolucci, S.}} \yr{1990}  \at{Direct numerical simulation of
  two-dimensional turbulent natural convection in an enclosed cavity}.
  \jt{J.~Fluid Mech.}  \bvol{215},  \pg{229--262}.

\bibitem[Sharp(1983)]{sharp1983overview}
{\sc \au{Sharp, D.~H.}} \yr{1983}  \bt{Overview of rayleigh-taylor
  instability}. {\em Tech. Rep.\/}.  \org{Los Alamos National Lab., NM (USA)}.

\bibitem[Shishkina(2016)]{shishkina2016momentum}
{\sc \au{Shishkina, O.}} \yr{2016}  \at{Momentum and heat transport scalings in
  laminar vertical convection}.  \jt{Phys.~Rev.~E}  \bvol{93},  \pg{051102}.

\bibitem[Spiegel \& Veronis(1960)]{spiegel1960boussinesq}
{\sc \au{Spiegel, E.~A.} \& \au{Veronis, G.}} \yr{1960}  \at{On the boussinesq
  approximation for a compressible fluid}.  \jt{Astrophy. J.}  \bvol{131},
  \pg{442}.

\bibitem[de~Vahl~Davis \& Jones(1983)]{de1983natural}
{\sc \au{de~Vahl~Davis, G.} \& \au{Jones, I.~P.}} \yr{1983}  \at{Natural
  convection in a square cavity: a comparison exercise}.
  \jt{Int.~J.~Numer.~Meth.~Fl.}  \bvol{3},  \pg{227--248}.

\bibitem[Wang {\em et~al.\/}(2019)Wang, Xia, Yan, Sun \& Wan]{wang2019non}
{\sc \au{Wang, Q.}, \au{Xia, S.-N.}, \au{Yan, R.}, \au{Sun, D.-J.} \& \au{Wan,
  Z.-H.}} \yr{2019}  \at{Non-oberbeck-boussinesq effects due to large
  temperature differences in a differentially heated square cavity filled with
  air}.  \jt{Int.~J.~Heat~Mass~Trans.}  \bvol{128},  \pg{479--491}.

\bibitem[Xia {\em et~al.\/}(2016)Xia, W., Liu, Wang \& Sun]{xia2016flow}
{\sc \au{Xia, S.-N.}, \au{W., Z.-H.}, \au{Liu, S.}, \au{Wang, Q.} \& \au{Sun,
  D.-J.}} \yr{2016}  \at{Flow reversals in {Rayleigh--B{\'e}nard} convection
  with {non-Oberbeck--Boussinesq effects}}.  \jt{J.~Fluid Mech.}  \bvol{798},
  \pg{628--642}.

\bibitem[Yang {\em et~al.\/}(2016)Yang, Verzicco \& Lohse]{yang2016vertically}
{\sc \au{Yang, Y.}, \au{Verzicco, R.} \& \au{Lohse, D.}} \yr{2016}
  \at{Vertically bounded double diffusive convection in the finger regime:
  Comparing no-slip versus free-slip boundary conditions}.
  \jt{Phys.~Rev.~Lett.}  \bvol{117}~(18),  \pg{184501}.

\bibitem[Zhang {\em et~al.\/}(2020)Zhang, Xia, Zhou \&
  Chen]{zhang2020controlling}
{\sc \au{Zhang, S.}, \au{Xia, Z.}, \au{Zhou, Q.} \& \au{Chen, S.}} \yr{2020}
  \at{Controlling flow reversal in two-dimensional {Rayleigh-B\'{e}nard}
  convection}.  \jt{J.~Fluid Mech.}  \bvol{891},  \pg{R4}.

\bibitem[Zhang \& Bao(2015)]{zhang2015direct}
{\sc \au{Zhang, Y.-Z.} \& \au{Bao, Y.}} \yr{2015}  \at{Direct solution method
  of efficient large-scale parallel computation for 3d turbulent
  {Rayleigh-Benard convection}}.  \jt{Acta~Phys.~Sin.}  \bvol{64},
  \pg{154702}.

\end{thebibliography}

\clearpage

\end{document}